\documentstyle[12pt]{article}

\def\hs{Haldane--Shastry}
\def\cs{Calogero--Sutherland}
\def\SUN{$SU(n)$}
\def\ket[#1]{\:\vert\,{#1}\,\rangle}
\def\notin{{\in}\kern-0.4em{/}\kern+0.05em}
\begin{document}

\begin{titlepage}

\def\thefootnote{\fnsymbol{footnote}}
\rightline{March 1993}
\rightline{cond-mat/9303050}
\vspace*{\fill}

\begin{center}
{\Large Spectrum of a spin chain with inverse square exchange} \\
\vfill
{\sc Holger Frahm\vspace{1.5 em}}\footnote{%
e-mail: {\tt frahm@kastor.itp.uni-hannover.de}}\\
{\sl Physikalisches Institut, Universit\"at Bayreuth, D-8580~Bayreuth,
Germany} \\ and \\
{\sl Institut f\"ur Theoretische Physik, Universit\"at Hannover,
D-3000~Hannover~1, Germany}\footnote{permanent address} \\
\vfill
ABSTRACT
\end{center}
\setlength{\baselineskip}{13pt}
The spectrum of a one-dimensional chain of \SUN\ spins positioned at
the static equilibrium positions of the particles in a corresponding
classical Calogero system with an exchange interaction inversely
proportional to the square of their distance is studied. As in the
translationally invariant \hs\ model the spectrum is found to exhibit
a very simple structure containing highly degenerate
``super-multiplets''. The algebra underlying this structure is
identified and several sets of raising and lowering operators are
given explicitely. On the basis of this algebra and numerical studies
we give the complete spectrum and thermodynamics of the $SU(2)$ system.

\vfill
\centerline{Submitted to {\em Journal of Physics A}}

\vfill
\setcounter{footnote}{0}
\end{titlepage}

Since it was first introduced \cite{hald:88,shas:88}, the \hs\ model
for spin chains with inverse square exchange and its generalizations
to \SUN\ spins have attracted considerable interest
\cite{inoz:89}--\nocite{hald:91,kiak:92,kawa:92ab,shas:92,forr:92,%
halx:92,fomi:cm92}\cite{shsu:pp92}. The Hamiltonian of these systems
is given by the expression
\begin{equation}
  {\cal H} = \sum_{j<k} h_{jk} P_{jk}
  \label{eq:hh}
\end{equation}
where $h_{jk} = \sin^{-2}\left( {\pi\over L}(x_j-x_k) \right)$ is the
exchange constant for spins at lattice sites with coordinate $x_j$ and
$x_k$ and $P_{jk}$ is the operator that exchanges the spins at these
sites.  The $x_j$ are chosen to be equally spaced around a ring of
circumfence $L$. Haldane and Shastry have given the wavefunctions for
the antiferromagnetic ground state which is of product form---just as
in the \cs\ model of particles on a line interacting with inverse
square potential---and also found the possible energy levels for the
system. The spectrum of this model exhibits a very simple structure
including a highly degenerate ``super-multiplet'' structure. Very
recently, this structure has been shown to be the consequence of a
Yangian symmetry \cite{drin:85,cher:90} present in this model
\cite{halx:92}. This symmetry which is intimately related to the
integrability of a system may eventually provide a way to construct a
generating function for the conserved quantities \cite{berx:ht93},
equivalent to the transfer matrix in the nearest neighbour exchange
models that are soluble by the quantum inverse scattering method
\cite{fata:84}, however, this has not been acchieved yet.

A complete set of these conserved quantities for the \hs\ \SUN\ chain
has been found by Fowler and Minahan \cite{fomi:cm92}. Using the
exchange operator formalism \cite{poly:92}, developed recently to
study the integrability of the \cs\ models (see e.g.\ Refs.\
\cite{calo:69ab}--\nocite{calo:71,calo:75,suth:71ab,suth:71c}%
\cite{suth:72}), they succeeded in constructing operators commuting
with ${\cal H}$ provided that the lattice sites $x_j$ are equally
spaced.

This work has been generalized by Polychronakos \cite{poly:ht92} to
construct a new integrable model of a spin chain with exchange
interaction related to another one of the classical \cs\ models. In a
calculation completely analogeous to that of Fowler and Minahan, he
was able to construct a complete set of invariants for the Hamiltonian
(\ref{eq:hh}) with exchange coupling
\begin{equation}
   h_{jk}={1 \over{(x_j-x_k)^2}}
   \label{oox2}
\end{equation}
provided the $x_j$ are the static equilibrium positions of particles
in the classical $N$-body Calogero system with potential
\cite{calo:69ab,calo:71}
\begin{equation}
  V(x_1,\ldots,x_N) = {1\over 2} \sum_j x_j^2
                    + \sum_{j<k} {1\over{(x_j-x_k)^2}},
\end{equation}
namely
\begin{equation}
  x_j = \sum_{k\ne j} {2\over(x_j-x_k)^3}.
  \label{inv3}
\end{equation}
(The same requirement leads to the lattice of equally spaced $x_j$ in
the \hs\ model).

While this construction of complete sets of commuting integrals
provides a proof for the integrability of the models (\ref{eq:hh}) it
does not give any hint on how to construct eigenstates and thus study
the energy spectrum. This is in contrast to the model with nearest
neighbour exchange where the algebraic Bethe Ansatz provides the
complete solution of this problem \cite{fata:84}. For some of the \cs\
models \cite{suth:71ab,suth:72} and the $SU(2)$~\hs\ model
\cite{hald:88} it has been shown that an {\em asymptotic} Bethe Ansatz
reproduces the exact spectrum and thermodynamics. However, a rigorous
derivation of the corresponding equations is still lacking.

The purpose of this letter is to present results of a study of the
spectrum of the integrable spin chain (\ref{eq:hh}) with inverse
square exchange (\ref{oox2}). Raising and lowering operators
connecting states of different energy are constructed which explain
the simple spectrum of this model and allow to obtain explicit
expressions for some eigenstates for the $SU(2)$ chain---including the
antiferromagnetic ground state which is again of product form.
Finally, an effective Hamiltonian is given that reproduces the
complete spectrum and degeneracies of the spin chain Hamiltonian.
Similar as in \cite{geru:92} this allows to study the thermodynamics
of the system.

At first sight, the condition (\ref{inv3}) fixing the positions $x_j$
of spins appears to be a major obstacle in studying the spectrum of
this system with analytical methods. Fortunately, the solutions $x_j$
of this equations can be identified as being the $N$ roots of the
Hermite polynomial $H_N(x)$ of degree $N$. In fact, it is
straightforward to obtain Eqs.~(\ref{inv3}) from the differential
equation defining $H_N(x)$, namely $y''-2xy'+2Ny=0$ (see e.g.\
\cite{szeg:book}). More relations between the $x_j$ of the type
(\ref{inv3}) that are useful in computing certain properties of the
system can be found in an analogeous way, e.g.\
\begin{equation}
   x_j = \sum_{k\ne j} {1\over{x_j-x_k}} \; ,\qquad
   2(N-1)-x_j^2 = \sum_{k\ne j} {1\over(x_j-x_k)^2}.
   \label{inv12}
\end{equation}
For large $N$ the density of roots of $H_N$ and hence the density of
sites in this nonuniform lattice is known to be
\begin{equation}
  \rho_N(x) = {1\over \pi} \sqrt{2\,N+1-x^2}.
\end{equation}

To proceed, we choose a representation of the permutation operator in
(\ref{eq:hh}) in terms of \SUN\ spin-operators $J_i^\alpha$
defined to be the $n^2-1$ traceless Hermitian $n\times n$-matrices of
the fundamental representation of the Lie algebra $[J^\alpha,
J^\beta]=f^{\alpha \beta \gamma}J^\gamma$ normalized such that
${\rm{Tr}}(J^\alpha J^\beta) = {1\over2}\delta^{\alpha\beta}$. The
$f^{\alpha\beta\gamma}$ are the antisymmetric structure constants of
\SUN\ and $J_i$ is understood as acting in the space of the spin at
site $i$ only. For $n=2$, $f^{\alpha\beta\gamma} =
i\varepsilon^{\alpha\beta\gamma}$ and $2 J_i^\alpha$ are
Pauli-matrices.  Through these spin-operators the permutation operator
$P_{jk}$ can be expressed as
\begin{equation}
  P_{jk} = {1\over n}\; {\bf 1}_j \otimes {\bf 1}_k
         + 2 \sum_\alpha J_j^\alpha \otimes J_k^\alpha.
  \label{permutop}
\end{equation}
(${\bf 1}_j$ is the $n\times n$ unit matrix acting in the Hilbert
space of the spin at site $j$.) From this one has to expect the
eigenvalues of (\ref{eq:hh}) to be grouped into \SUN\ multiplets
since
\begin{equation}
   \left[{\cal H},Q_0^\alpha\right] = 0,
   \qquad \alpha = 1,\ldots,n^2-1
\end{equation}
with $Q_0^\alpha = \sum_j J_j^\alpha$ being the components of the
total spin.

However, numerical diagonalization of the Hamiltonian ${\cal H}$ for
$SU(2)$ spins with exchange coupling (\ref{oox2}) on small lattices
($N \le 10$) satisfying (\ref{inv3}) shows---very similar to the
observations in the \hs\ model---a much simpler structure (see
Fig.~\ref{fig:spectrum}): In addition to the expected $SU(2)$
multiplets one finds that all the energies are integers, and that
there are additional degeneracies between states having different
total spin.

Similar ``super-multiplets'' have been observed in the periodic
$SU(2)$ \hs\ model \cite{hald:88,shas:88}. The generators of the
algebra corresponding to this symmetry have been identified in
\cite{halx:92}. Following this work, we are led to study the
commutators of the Hamiltonian (\ref{eq:hh}) with the operators
\begin{equation}
  L_1^\alpha = \sum_{j\ne k} w_{jk}
     f^{\alpha\beta\gamma} J_j^\beta J_k^\gamma.
   \label{eq:l1def}
\end{equation}
It turns out that choosing $w_{jk}=1/(x_j-x_k)$ leads to the vanishing
of terms containing spin-operators at three different sites.
Furthermore, since the $J_j^\alpha$ act in the fundamental
representation, one finds using (\ref{inv3}) that\footnote{%%%%
%%%%
For the \SUN\ \hs\ model an operator of the form (\ref{eq:l1def}) has
been shown to commute with ${\cal H}$ \cite{halx:92}.}
%%%%
\begin{equation}
  \left[ {\cal H},L_1^\alpha \right] =
  L_0^\alpha \equiv \sum_{j=1}^N x_j J_j^\alpha.
\end{equation}
Similarly, one obtains $\left[{\cal H},L_0^\alpha\right]=L_1^\alpha$
from which $n^2-1$ pairs of raising (lowering) operators can be
defined with
\begin{equation}
   Q^\alpha_\pm=L_0^\alpha \pm L_1^\alpha.
   \label{eq:rl}
\end{equation}
It is easily checked that
\begin{equation}
   \left[Q_0^\alpha,L_i^\beta\right]
   =f^{\alpha\beta\gamma}L_i^\gamma, \qquad i=0,1.
\end{equation}
By construction an equivalent relation holds for the commutator
$\left[Q^\alpha_0,Q_\pm^\beta\right]$.

This observation allows to study the spectrum of the Hamiltonian
(\ref{eq:hh}) in more detail: To be specific we consider the
case of $SU(2)$ corresponding to an isotropic spin-${1\over 2}$ chain
in the following. From $Q^z_+\ket[0;m]=0$ one easily finds for the
state of highest energy with $Q_0^z\ket[0;m] = \left( {1\over2}N-m
\right) \ket[0;m]$ ($\ket[0] = |\uparrow_1\cdots\uparrow_N\rangle$ is
the ferromagnetic vacuum):
\begin{equation}
   \ket[0;m] \propto \sum_{j_1<j_2<\ldots <j_m} \psi_{j_1 j_2 \ldots j_m}
                 \prod_{s=1}^m J_{j_s}^- \ket[0]
   \label{eq:fmstates}
\end{equation}
($J_j^\pm=J_j^x\pm i J_j^y$) with $\psi_{j_1 \ldots j_m} \equiv 1$.
This is the multiplet with total spin $S=N/2$, i.e.\ all ferromagnetic
states $\ket[0;m]=\left(Q_0^-\right)^m \ket[0]$ in the system.  Acting
on $\ket[0;m]$ with $Q_-^z$ and $Q_-^\pm=Q_-^x\pm iQ_-^y$ eigenstates
with lower energies can be constructed. For example, the states with a
single overturned spin as compared to the ferromagnetic vacuum
$\ket[0]$, namely $\left(Q_-^z\right)^n \ket[0;1]$, $n=0,\ldots N-1$
are of the form (\ref{eq:fmstates}) with
\begin{equation}
   \psi_j^{(n)} = \pi^{(n)}(x_j).  \label{eq:1mag}
\end{equation}
Here $\pi^{(n)}(x)$ is a polynomial in $x$ of degree $n$, belonging to
the set of orthogonal polynomials on the discrete set
$\left\{x_j\right\}$ with unit weight function defined through the
recursion relations
\begin{equation}
  \pi^{(n)}(x) = A_n \, x \, \pi^{(n-1)}(x) \,-\, C_n \, \pi^{(n-2)}(x),
  \quad n=1,\ldots, N-1
\end{equation}
with $\pi^{(-1)}\equiv 0$, $\pi^{(0)}(x) \equiv 1/\sqrt{N}$ and
\begin{equation}
  A_n = \sqrt{2\over{N-n}}, \qquad C_n = {A_n \over A_{n-1}}.
\end{equation}
Similarly, it can be shown, that each state with energy $-n$ as
compared to the ferromagnetic state and magnetization ${1\over2}N-m$
can be written as a polynomial in $m$ variables $y_j \in
\left\{x_k\right\}$, $j=1,\ldots,m$
where the powers in the terms $\prod_j (y_j)^{m_j}$ satisfy $\sum_j
m_j \le n$.

Of course, this form of the eigenstates is not very useful when one is
interested in studying the properties of the antiferromagnetic ground
state with $m={1\over 2}N$ and energy $E_0=-({1\over 2}N)^2$ (for even
number of lattice sites $N$). For the \hs\ model it is known that the
ground state can be written in product form. The same turns out to be
true for this model. One finds that the lowest state with total
magnetization $N/2-m$ can be written in the form (\ref{eq:fmstates})
with
\begin{equation}
  \psi_{j_1 \ldots j_m} =
	\prod_{j     \in \{j_\alpha\}}  \left(-1\right)^j
        \prod_{j,k \in \{j_\alpha\}}    \left(x_j - x_k\right)
        \prod_{j,k \notin \{j_\alpha\}} \left(x_j - x_k\right).
\end{equation}
Higher energy states can be obtained through the action of $Q^\alpha_+$.

It remains to be shown that the {\em complete} spectrum is obtained
under the action of the operators $Q_0^\alpha$, $Q_\pm^\alpha$ on the
ferromagnetic vacuum $\ket[0]$. It is conceivable that successive
application of different sets of these operators leads to the same
final state. For a complete answer to this question the algebra of the
operators $Q_0^\alpha$, $Q_\pm^\alpha$ needs further study. However, a
simple example shows that in general different states are generated by
different operator sequences in the $Q_0^z$--$E$ plane
(Fig.~\ref{fig:spectrum}): starting with the one-magnon state
$\ket[\psi^{(1)}]$ (Eq.~(\ref{eq:1mag})) one obtains
\begin{equation}
   \ket[a\, ;2] = Q_0^- Q_-^z \ket[\psi^{(1)}] =Q_0^- \ket[\psi^{(2)}]
        \propto \sum_{j\ne k} \left(x_j^2+x_k^2 -(N-1)\right)
        J_j^- J_k^- \ket[0]
\end{equation}
A different state with the same energy and magnetization can be
constructed from $\ket[\psi^{(1)}]$ through the action of $Q_-^-$
resulting in
\begin{equation}
   \ket[b\, ;2] = \left( \sum_j x_j J_j^-
       + 2 \sum_{j\ne k} w_{jk} J_j^z J_k^- \right)\ket[\psi^{(1)}]
   = \sum_{j\ne k} \left( 2\, x_j x_k +1 \right) J_j^- J_k^- \ket[0]
\end{equation}
It is straightforward to show that $\ket[a\, ;2]$ and $\ket[b\, ;2]$
are indeed linearly independent. Note that they are {\em not}
orthogonal though.

Finally, we observe that the simple structure of the spectrum found
from the numerical investigation of finite chains allows for an
alternative description through an effective single particle
Hamiltonian with additional energy proportional to the square of the
total number of particles
\begin{equation}
  {\cal H}^{\rm eff} = \sum_{k=1}^N \left(1-k\right) n_k
                     - \sum_{k<k'} n_k n_{k'}
                     + \left( {N\over2} \right)^2.
  \label{eq:heff}
\end{equation}
The occupation numbers $n_k$ take values $0,1$. From (\ref{eq:heff})
an expression for the partition function of the spin chain can be
obtained:
\begin{equation}
  {\cal Z} = \sum_{k=0}^{N} z^{\left(N/2-k\right)^2}
             \prod_{r=1}^{k} {{1-z^{N-r+1}}\over{1-z^r}}, \qquad
  z = {\rm e}^{-\beta}
  \label{eq:zzz}
\end{equation}
with the leading terms for low temperatures (and sufficiently large
$N$)
\begin{equation}
  {\cal Z} = 1+3\,z+4\,z^{2}+7\,z^{3}+13\,z^{4}+O(z^{5})
\end{equation}
To perform the thermodynamic limit ($N\to\infty$) in a meaningful way
the energies have to be rescaled by a factor of $N$. This makes the
excitations in the infinite system massless, hence the model has a
critical point at $T=0$. In this limit Eq.~(\ref{eq:zzz}) can be
brought into a closed form. The resulting expression for the free
energy per spin reads
\begin{equation}
   {{\cal F}\over N} = - T^2 \left( {\pi^2 \over 6}
                    + 2 f\left(1+{\rm e}^{-\beta/2} \right) \right).
   \label{eq:finf}
\end{equation}
where $f(x)$ is the dilogarithm
\begin{equation}
   f(x) =  -\int_x^1 {\rm d}t {{\ln t}\over{1-t}}.
   \label{eq:dilog}
\end{equation}
The asymptotic behaviour of ${\cal F}$ at low and high temperatures is
given by
\begin{eqnarray}
   {{\cal F}\over N} &\sim& -T^2 \left( {\pi^2\over 6}
                     - 2\,{\rm e}^{-\beta/2} +o({\rm e}^{-\beta}) \right)
                    \qquad {\rm for~} T\to 0 \nonumber \\
                     &\sim& -T \ln 2 + {1\over 8} + o(T^{-1})
                    \qquad {\rm for~} T\to\infty.
\end{eqnarray}
{}From (\ref{eq:finf}) further thermodynamic quantities are easily
extracted.

In this letter I have studied a new integrable model in the class of
spin chains related to the classical \cs\ models. As in other lattice
models with inverse square exchange together with the integrability
one finds a very simple and highly degenerate spectrum. In the model
investigated here this can be understood in terms of the existence of
an algebra of raising and lowering operators (\ref{eq:rl}). It remains
to be studied whether these operators can be interpreted in a way
similar to the corresponding ones in the periodic \hs\ spin chain,
which have been shown to be the level-1 generators of a Yangian. A
detailed understanding of this algebra is also necessary to prove the
equivalence of the original Hamiltonian (\ref{eq:hh}) and the
effective one (\ref{eq:heff}) and may lead to a rigourous foundation
for the asymptotic Bethe Ansatz in the integrable models with inverse
square exchange.

The author thanks V.~Rittenberg and H.~R\"oder for useful discussions.

\newpage
\centerline{\large \bf Figure Captions}

\begin{figure}[h]
\caption{\label{fig:spectrum}}
Spectrum of the spin chain (\ref{eq:hh}) with exchange
(\ref{oox2}) for $N=6$ spins (energy vs.\ $z$-component of total
spin $Q_0^z$). The degeneracies are given in parentheses.

\vspace{2 cm}

\setlength{\unitlength}{0.012500in}%
\begin{picture}(336,348)(-49,450)
\thicklines
\put( 60,783){\circle*{4}}
\put(105,783){\circle*{4}}
\put(105,753){\circle*{4}}
\put(105,723){\circle*{4}}
\put(105,693){\circle*{4}}
\put(105,663){\circle*{4}}
\put(105,633){\circle*{4}}
\put(150,633){\circle*{4}}
\put(150,663){\circle*{4}}
\put(150,693){\circle*{4}}
\put(150,723){\circle*{4}}
\put(150,753){\circle*{4}}
\put(150,783){\circle*{4}}
\put(195,783){\circle*{4}}
\put(195,753){\circle*{4}}
\put(195,723){\circle*{4}}
\put(195,693){\circle*{4}}
\put(195,663){\circle*{4}}
\put(195,633){\circle*{4}}
\put(195,603){\circle*{4}}
\put(150,603){\circle*{4}}
\put(150,573){\circle*{4}}
\put(195,573){\circle*{4}}
\put(195,543){\circle*{4}}
\put(150,543){\circle*{4}}
\put(240,543){\circle*{4}}
\put(240,573){\circle*{4}}
\put(240,603){\circle*{4}}
\put(240,633){\circle*{4}}
\put(240,663){\circle*{4}}
\put(240,693){\circle*{4}}
\put(240,723){\circle*{4}}
\put(240,753){\circle*{4}}
\put(240,783){\circle*{4}}
\put(285,783){\circle*{4}}
\put(285,753){\circle*{4}}
\put(285,723){\circle*{4}}
\put(285,693){\circle*{4}}
\put(285,663){\circle*{4}}
\put(285,633){\circle*{4}}
\put(330,783){\circle*{4}}
\put( 30,525){\vector( 0, 1){240}}
\put( 75,480){\vector( 1, 0){240}}
\put(180,513){\line( 1, 0){ 30}}
\put( 45,783){\line( 1, 0){ 30}}
\put( 90,783){\line( 1, 0){ 30}}
\put( 90,753){\line( 1, 0){ 30}}
\put( 90,723){\line( 1, 0){ 30}}
\put( 90,693){\line( 1, 0){ 30}}
\put( 90,663){\line( 1, 0){ 30}}
\put( 90,633){\line( 1, 0){ 30}}
\put(135,633){\line( 1, 0){ 30}}
\put(135,663){\line( 1, 0){ 30}}
\put(135,693){\line( 1, 0){ 30}}
\put(135,723){\line( 1, 0){ 30}}
\put(135,753){\line( 1, 0){ 30}}
\put(135,783){\line( 1, 0){ 30}}
\put(180,783){\line( 1, 0){ 30}}
\put(180,753){\line( 1, 0){ 30}}
\put(180,723){\line( 1, 0){ 30}}
\put(180,693){\line( 1, 0){ 30}}
\put(180,663){\line( 1, 0){ 30}}
\put(180,633){\line( 1, 0){ 30}}
\put(180,603){\line( 1, 0){ 30}}
\put(195,513){\circle*{4}}
\put(135,603){\line( 1, 0){ 30}}
\put(199,576){\makebox(0,0)[lb]{\smash{\rm (2)}}}
\put(135,573){\line( 1, 0){ 30}}
\put(180,573){\line( 1, 0){ 30}}
\put(180,543){\line( 1, 0){ 30}}
\put(135,543){\line( 1, 0){ 30}}
\put(225,543){\line( 1, 0){ 30}}
\put(225,573){\line( 1, 0){ 30}}
\put(225,603){\line( 1, 0){ 30}}
\put(225,633){\line( 1, 0){ 30}}
\put(225,663){\line( 1, 0){ 30}}
\put(225,693){\line( 1, 0){ 30}}
\put(225,723){\line( 1, 0){ 30}}
\put(225,753){\line( 1, 0){ 30}}
\put(225,783){\line( 1, 0){ 30}}
\put(270,783){\line( 1, 0){ 30}}
\put(270,753){\line( 1, 0){ 30}}
\put(270,723){\line( 1, 0){ 30}}
\put(270,693){\line( 1, 0){ 30}}
\put(270,663){\line( 1, 0){ 30}}
\put(270,633){\line( 1, 0){ 30}}
\put(315,783){\line( 1, 0){ 30}}
\put(186,450){\makebox(0,0)[lb]{\smash{$Q_0^z$}}}
\put(  9,645){\makebox(0,0)[lb]{\smash{$E$}}}
\put(199,516){\makebox(0,0)[lb]{\smash{\rm (1)}}}
\put( 64,786){\makebox(0,0)[lb]{\smash{\rm (1)}}}
\put(109,786){\makebox(0,0)[lb]{\smash{\rm (1)}}}
\put(109,756){\makebox(0,0)[lb]{\smash{\rm (1)}}}
\put(109,726){\makebox(0,0)[lb]{\smash{\rm (1)}}}
\put(109,696){\makebox(0,0)[lb]{\smash{\rm (1)}}}
\put(109,666){\makebox(0,0)[lb]{\smash{\rm (1)}}}
\put(109,636){\makebox(0,0)[lb]{\smash{\rm (1)}}}
\put(154,756){\makebox(0,0)[lb]{\smash{\rm (1)}}}
\put(154,786){\makebox(0,0)[lb]{\smash{\rm (1)}}}
\put(199,786){\makebox(0,0)[lb]{\smash{\rm (1)}}}
\put(199,756){\makebox(0,0)[lb]{\smash{\rm (1)}}}
\put(154,576){\makebox(0,0)[lb]{\smash{\rm (1)}}}
\put(199,546){\makebox(0,0)[lb]{\smash{\rm (1)}}}
\put(154,546){\makebox(0,0)[lb]{\smash{\rm (1)}}}
\put(244,546){\makebox(0,0)[lb]{\smash{\rm (1)}}}
\put(244,576){\makebox(0,0)[lb]{\smash{\rm (1)}}}
\put(244,756){\makebox(0,0)[lb]{\smash{\rm (1)}}}
\put(244,786){\makebox(0,0)[lb]{\smash{\rm (1)}}}
\put(289,786){\makebox(0,0)[lb]{\smash{\rm (1)}}}
\put(289,756){\makebox(0,0)[lb]{\smash{\rm (1)}}}
\put(289,726){\makebox(0,0)[lb]{\smash{\rm (1)}}}
\put(289,696){\makebox(0,0)[lb]{\smash{\rm (1)}}}
\put(289,666){\makebox(0,0)[lb]{\smash{\rm (1)}}}
\put(289,636){\makebox(0,0)[lb]{\smash{\rm (1)}}}
\put(334,786){\makebox(0,0)[lb]{\smash{\rm (1)}}}
\put(154,726){\makebox(0,0)[lb]{\smash{\rm (2)}}}
\put(199,726){\makebox(0,0)[lb]{\smash{\rm (2)}}}
\put(244,726){\makebox(0,0)[lb]{\smash{\rm (2)}}}
\put(154,696){\makebox(0,0)[lb]{\smash{\rm (2)}}}
\put(154,666){\makebox(0,0)[lb]{\smash{\rm (3)}}}
\put(199,696){\makebox(0,0)[lb]{\smash{\rm (3)}}}
\put(244,696){\makebox(0,0)[lb]{\smash{\rm (2)}}}
\put(199,666){\makebox(0,0)[lb]{\smash{\rm (3)}}}
\put(244,666){\makebox(0,0)[lb]{\smash{\rm (3)}}}
\put(154,636){\makebox(0,0)[lb]{\smash{\rm (2)}}}
\put(244,636){\makebox(0,0)[lb]{\smash{\rm (2)}}}
\put(199,636){\makebox(0,0)[lb]{\smash{\rm (3)}}}
\put(154,606){\makebox(0,0)[lb]{\smash{\rm (2)}}}
\put(199,606){\makebox(0,0)[lb]{\smash{\rm (3)}}}
\put(244,606){\makebox(0,0)[lb]{\smash{\rm (2)}}}
\end{picture}

\end{figure}


\begin{thebibliography}{10}

\bibitem{hald:88}
F.~D. Haldane,
\newblock Phys. Rev. Lett. {\bf 60}, 635 (1988).

\bibitem{shas:88}
B.~S. Shastry,
\newblock Phys. Rev. Lett. {\bf 60}, 639 (1988).

\bibitem{inoz:89}
V.~I. Inozemtsev,
\newblock J. Stat. Phys. {\bf 59}, 1143 (1989).

\bibitem{hald:91}
F.~D.~M. Haldane,
\newblock Phys. Rev. Lett. {\bf 66}, 1529 (1991).

\bibitem{kiak:92}
H.~Kiwata and Y.~Akutsu,
\newblock J. Phys. Soc. Japan {\bf 61}, 1441 (1992).

\bibitem{kawa:92ab}
N.~Kawakami,
\newblock Phys. Rev. B {\bf 46}, 1005 and 3191 (1992).

\bibitem{shas:92}
B.~S. Shastry,
\newblock Phys. Rev. Lett. {\bf 69}, 164 (1992).

\bibitem{forr:92}
P.~J. Forrester,
\newblock J. Phys. A {\bf 25}, 5447 (1992).

\bibitem{halx:92}
F.~D.~M. Haldane, Z.~N.~C. Ha, J.~C. Talstra, D.~Bernard, and V.~Pasquier,
\newblock Phys. Rev. Lett. {\bf 69}, 2021 (1992).

\bibitem{fomi:cm92}
M.~Fowler and J.~A. Minahan,
\newblock U of Virginia preprint UVA-HET-92-07 (1992).

\bibitem{shsu:pp92}
B.~S. Shastry and B.~Sutherland,
\newblock preprint (1992).

\bibitem{drin:85}
V.~G. Drinfel'd,
\newblock Soviet Math. Dokl. {\bf 32}, 254 (1985).

\bibitem{cher:90}
I.~Cherednik,
\newblock {\em Notes on Affine Hecke Algebras {I.}},
\newblock preprint BONN-HE-90-04, U Bonn, (1990).

\bibitem{berx:ht93}
D.~Bernard, M.~Gaudin, F.~D.~M. Haldane, and V.~Pasquier,
\newblock Saclay preprint SPhT-93-006 (1993).

\bibitem{fata:84}
L.~A. Takhtajan and L.~Faddeev,
\newblock J. Sov. Math. {\bf 24}, 241 (1984),
\newblock [Zap. Nauch. Semin. LOMI {\bf 109}, 134 (1981)].

\bibitem{poly:92}
A.~P. Polychronakos,
\newblock Phys. Rev. Lett. {\bf 69}, 703 (1992).

\bibitem{calo:69ab}
F.~Calogero,
\newblock J. Math. Phys. {\bf 10}, 2191 and 2197 (1969).

\bibitem{calo:71}
F.~Calogero,
\newblock J. Math. Phys. {\bf 12}, 419 (1971).

\bibitem{calo:75}
F.~Calogero,
\newblock Lett. Nuovo Cimento {\bf 13}, 507 (1975).

\bibitem{suth:71ab}
B.~Sutherland,
\newblock J. Math. Phys. {\bf 12}, 246 and 251 (1971).

\bibitem{suth:71c}
B.~Sutherland,
\newblock Phys. Rev. A {\bf 4}, 2019 (1971).

\bibitem{suth:72}
B.~Sutherland,
\newblock Phys. Rev. A {\bf 5}, 1372 (1972).

\bibitem{poly:ht92}
A.~P. Polychronakos,
\newblock preprint {NTUA} 34/92 (1992).

\bibitem{geru:92}
F.~Gebhard and A.~E. Ruckenstein,
\newblock Phys. Rev. Lett. {\bf 68}, 244 (1992).

\bibitem{szeg:book}
G.~Szeg{\H{o}},
\newblock {\em Orthogonal polynomials}, 4th edition,
\newblock Amer. Math. Soc., Providence (RI), 1975.

\end{thebibliography}
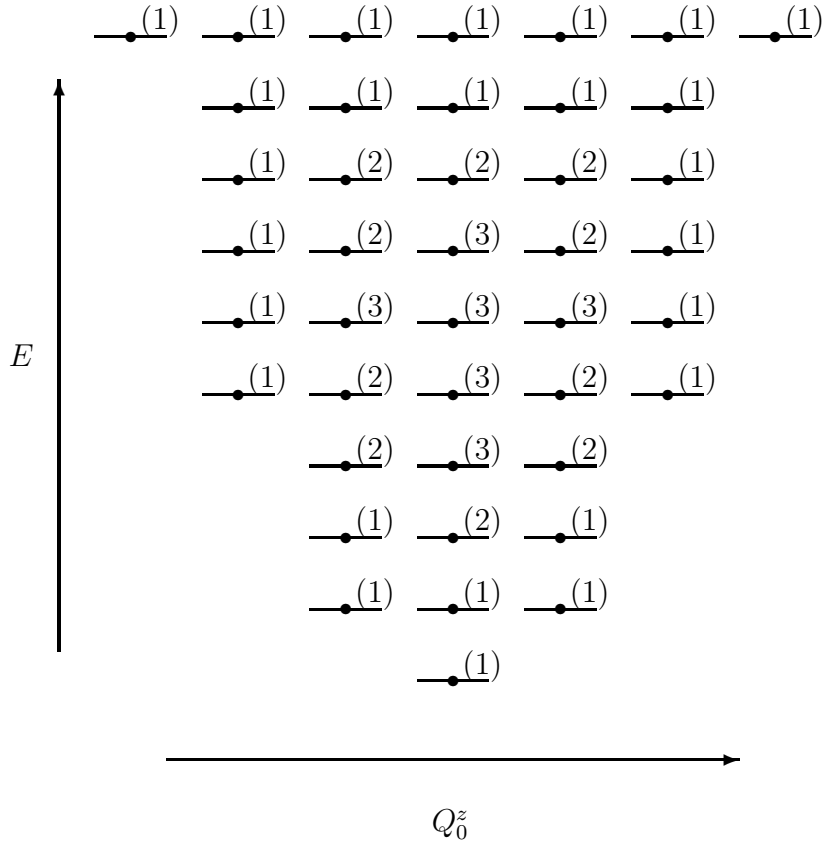
\end{document}